\documentclass[jcp,aps,showpacs,supergroupedaddress,epsfig,amsmath,amssymb,eqsecnum]{revtex4}
\usepackage{epsfig,amsmath,amsthm,amssymb,bm,epsf,graphics,psfrag,verbatim,subfigure,framed}
\usepackage[makeroom]{cancel}
\usepackage{mathtools}
\usepackage[usenames,dvipsnames]{color}
\usepackage{empheq}
\usepackage{bbm}
\usepackage{mathrsfs}
\usepackage{amsfonts}
\usepackage{color} 
\usepackage{pbox}
\usepackage{xcolor} 
\usepackage{multirow}
\newcommand{\be}{\begin{equation}}
\newcommand{\ee}{\end{equation}}
\newcommand{\ba}{\begin{eqnarray}}
\newcommand{\ea}{\end{eqnarray}}
\newcommand{\bw}{\begin{widetext}}
\newcommand{\ew}{\end{widetext}}

\newcommand{\rv}{{\mathbf{r}}}

\begin{document}

\title{Casimir-Polder interaction between an atom and a Chern insulator: 
\\
topological signature and long-range repulsion}
\author{Bing-Sui Lu}
\email{binghermes@gmail.com, bslu@siit.tu.ac.th}
\affiliation{Sirindhorn International Institute of Technology, Thammasat University, 
Pathum Thani 12120, Thailand}

\date{\today}


\begin{abstract}
We consider the Casimir-Polder interaction between a two-level atomic system and a Chern insulator for both the resonant and nonresonant channels. For a right circularly polarized excited atomic state near a Chern insulator with a negative Chern number $C$, the resonant Casimir-Polder force can be monotonically repulsive over a large range of separations. In the presence of the same Chern insulator, a right circularly polarized metastable atomic state is expected to experience a repulsive nonresonant Casimir-Polder force over a certain range of atom-surface separations in the far-field region. At still greater separations, the nonresonant Casimir-Polder force is expected to become attractive and exhibit a topological signature, being proportional to $(C\alpha)^2/(1+(C\alpha)^2)$, where $\alpha$ is the fine-structure constant. 
\end{abstract}

\maketitle

\section{Introduction}

A Chern insulator~\cite{qwz2006,liu2016,zhang2016,weng2015,asboth2016} is a two-dimensional topological insulator~\cite{tkachov,lu2021,lu2018} that has a nonzero Hall conductivity, with the zero-frequency Hall conductance becoming quantized as $C e^2/h$ (where $C$ is an integer known as the Chern number) as the temperature goes to zero. This quantization arises purely as a result of topology: in the zero-temperature and zero-frequency limit, the Kubo formula for the Hall conductance becomes the TKNN formula~\cite{tknn}, involving a topological invariant ($C$) which expresses the number of times a unit Bloch vector winds around a unit sphere as the momentum vector traverses the entire Brillouin zone~\cite{dubrovin}. It has been predicted that such a topological signature can be detected in the far-field behavior of the Casimir-Lifshitz force~\cite{bordag,milton} between two Chern insulators~\cite{pablo2014}. Furthermore, the surface magnetization that gives rise to a nonzero Hall conductivity makes the Chern insulator a nonreciprocal material~\cite{weng2015}. Being nonreciprocal, a pair of Chern insulators can experience a repulsive Casimir-Lifshitz force if their Chern numbers have opposite signs. Chern insulators have been fabricated by depositing a thin film of topological insulator material (such as bismuth telluride) on a substrate, doping the material with transition metal atoms (such as chromium or vanadium) and lowering the temperature to achieve a surface ferromagnetic phase transition. The experimental realizability of Chern insulators makes it timely to study the physics of their Casimir interaction. 

\

In this paper, we consider the Casimir-Polder interaction~\cite{CP-paper,ducloy-review,laliotis2021} between an atom in a circularly polarized state and a Chern insulator for both the resonant and nonresonant channels. 
In Sections~\ref{sec:CPshift} and \ref{sec:CPshift-res}, we report some of the key results from Refs.~\cite{lu2022} and \cite{lu2020}, while in Section~\ref{sec:CPshift-nonres}, we additionally report some unpublished results of our investigations. 
Because of the very low temperatures required to achieve surface magnetization in practice, we have restricted our consideration to the Casimir-Polder interaction at zero temperature. We shall see that analogous to the far-field Casimir-Lifshitz force behavior between two Chern insulators, the far-field limit of the nonresonant Casimir-Polder force between a metastable atom in a circularly polarized state and a Chern insulator can also exhibit a topological signature of the Chern insulator. Here, we take the far-field region to refer to those separations over which retardation effects are non-negligible, and the near-field region to refer to those separations over which retardation effects can be neglected. Furthermore, though the nonresonant Casimir-Polder force on the metastable atom in the extreme far-field limit is always attractive, there is a range of far-field separations over which the force can become repulsive. We shall also see that the resonant Casimir-Polder force on a circularly polarized excited atom can be repulsive over a large range of separations.  

\section{The Casimir-Polder energy shift}
\label{sec:CPshift}

An atom in free space isolated from radiation has energy levels that are given by the solution to Schroedinger's equation with the potential term being the Coulomb energy between the electron(s) and the nucleus~\cite{griffiths-QM}. Turning on the electric dipole interaction between the atom and radiation gives rise to shifts in the energy levels~\cite{bethe}. Additional shifts in the energy levels are generated if the atom is brought close to a surface, with the additional shifts being dependent on the atom-surface separation distance~\cite{wylie-sipe1,wylie-sipe2}. These additional, position-dependent energy level shifts are the Casimir-Polder energy shifts, and their position dependence leads to the existence of the Casimir-Polder force. The Casimir-Polder shift has a nonresonant contribution, which corresponds to the energy of polarizing the atom by vacuum fluctuations. For an excited atomic state, there can be a resonant contribution as well. Physically, the resonant contribution stems from the de-excitation of the atom which involves the emission of a real photon that then gets reflected by the surface back to the position of the atom. The resonant Casimir-Polder shift can be interpreted classically as the energy of interaction between a dipole and the image dipole~\cite{wylie-sipe2}. 

\

Let us consider an atom which is sufficiently small so that it can be approximated by a point dipole. The Hamiltonian expressing the interaction of an electric dipole ${\bm{\mu}}$ with the radiation field $\mathcal{E}$ is 
\be
H_{{\rm int}} = - {\bm{\mu}} \cdot \mathcal{E}. 
\ee
We adopt a simplified model of the atom as a two-level system. This means that we are restricting our consideration to a single (electric dipole) transition between two states with different energies in an atom within a certain timeframe. We also assume that the atom is sufficiently far from the Chern insulator, so that the atom's wavefunction does not overlap significantly with that of the insulator. Finally, as the radiation field is a weak field, we can assume linear response, whereby the field response is linearly related to the dipole source via an electromagnetic Green tensor.  

\

\begin{figure}[h]
    \centering
      \includegraphics[width=0.77\textwidth]{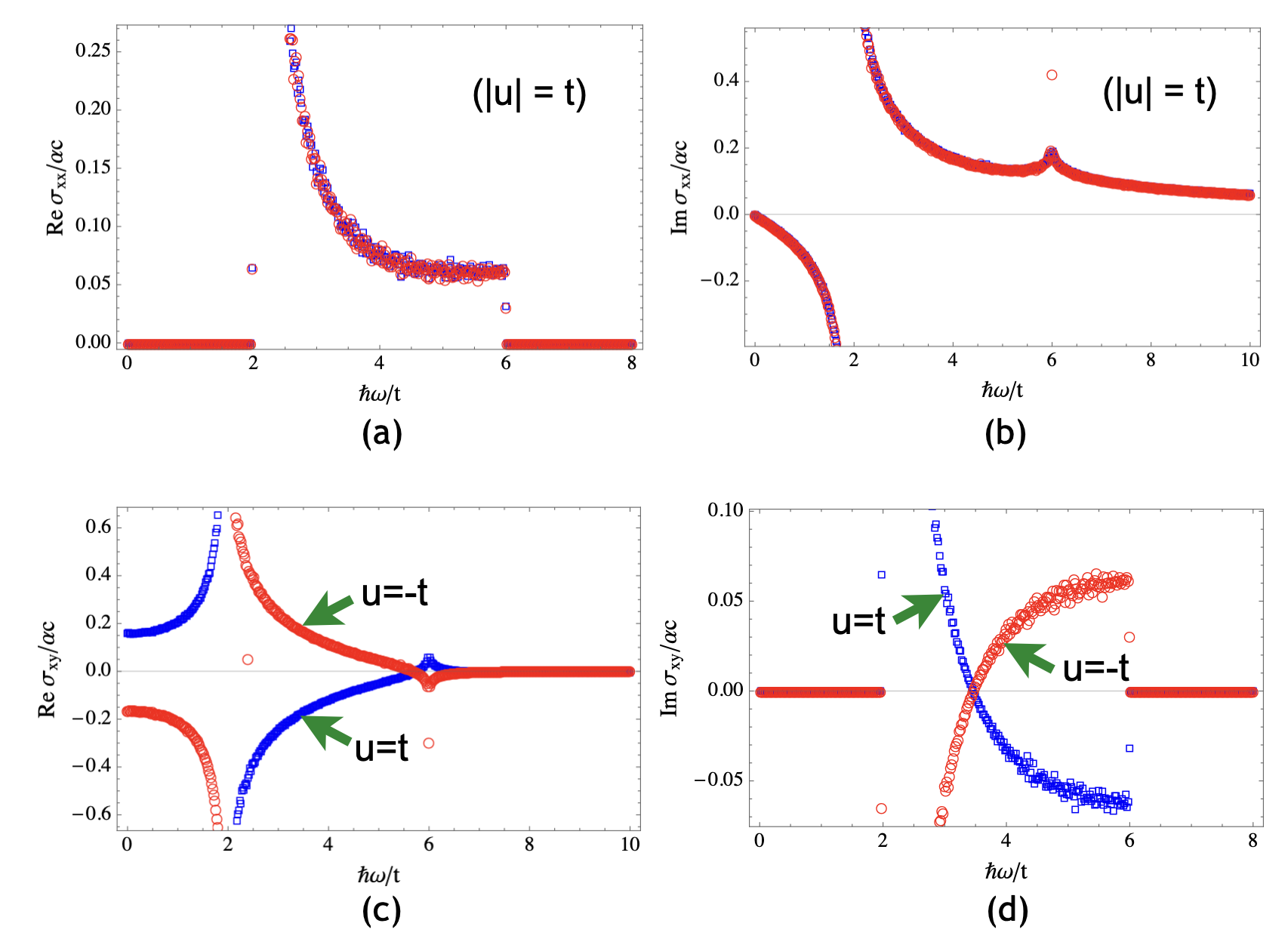}
      \caption{Frequency-dependent behavior of (a)~the real part of the longitudinal conductivity, (b)~the imaginary part of the longitudinal conductivity, (c)~the real part of the Hall conductivity, and (d)~the imaginary part of the Hall conductivity. The conductivity is calculated using Kubo's formula with the Qi-Wu-Zhang model Hamiltonian, Equation~(\ref{eq:QWZ}). The behavior for $u = t$ ($u = - t$) is shown by the squares (circles).} 
     \label{dispersion}
  \end{figure}

Based on the above assumptions, the Casimir-Polder shifts of the upper- and lower-energy states have been calculated for an atom near a Chern insulator~\cite{lu2022}. We denote the upper- and lower-energy states by $|1\rangle$ and $|0\rangle$ respectively. The Casimir-Polder shift for the upper-energy state, $\delta E_1^{CP}$, was found to be 
\be
\delta E_1^{CP} = \delta E_1^{{\rm vdw}} + \delta E_1^{{\rm cl}}, 
\ee
where $\delta E_1^{{\rm vdw}}$ is the nonresonant contribution, given by 
\be
\delta E_1^{{\rm vdw}} = \frac{1}{\pi} \mu_a^{10} \mu_b^{01}  \,
\int_0^\infty \!\!\!\! d\xi \, 
\Bigg(
\frac{\omega_{10} \big( G^R_{ab}({\bf r}_0,{\bf r}_0; i\xi) + G_{ba}^{R}({\bf r}_0,{\bf r}_0; i\xi) \big)}{2(\omega_{10}^2 + \xi^2)}
-
\frac{\xi \big( G^R_{ab}({\bf r}_0,{\bf r}_0; i\xi) - G_{ba}^{R}({\bf r}_0,{\bf r}_0; i\xi) \big)}{2i(\omega_{10}^2 + \xi^2)}
\Bigg). 
\ee
Here, $\rv_0$ is the position vector of the atom, $i\xi$ is an imaginary frequency, $a=x,y,z$ denotes the Cartesian component, ${\bm{\mu}}^{mn} \equiv \langle m | {\bm{\mu}} | n \rangle$ is the atomic dipole transition matrix element from $|n\rangle$ to $|m\rangle$, $\omega_{mn} \equiv (E_m - E_n)/\hbar$ is the energy difference between the unperturbed states divided by the Planck constant, and $G_{ab}^R$ is the reflection electromagnetic Green tensor. The Green tensor components relevant to our purpose are $G_{xx}^R$ and $G_{xy}^R$, which are given by~\cite{lu2022}
\ba
G^R_{xx}({\bf r}_0, {\bf r}_0, \omega) 
&=& 
G^R_{yy}({\bf r}_0, {\bf r}_0, \omega) 
=
\frac{i}{2} \Big( \frac{\omega}{c} \Big)^2 \! 
\int_{0}^\infty \!\!\!\! dk_\parallel \frac{k_\parallel}{k_z} \, e^{2 i k_z z} 
\big( 
r_{ss} - \frac{c^2 k_z^2}{\omega^2} r_{pp} 
\big), 
\nonumber\\
G^R_{xy}({\bf r}_0, {\bf r}_0, \omega) 
&=& 
- G^R_{yx}({\bf r}_0, {\bf r}_0, \omega) 
=
\frac{i}{2} \Big( \frac{\omega}{c} \Big) \! 
\int_{0}^\infty \!\!\!\! dk_\parallel k_\parallel e^{2 i k_z z} 
\left( r_{ps} + r_{sp} \right). 
\label{GR0}
\ea
Here, $\omega$ is the frequency, $k_\parallel = \sqrt{k_x^2+k_y^2}$ is the wavevector in the plane of the Chern insulator, $k_z = \sqrt{\omega^2/c^2 - k_\parallel^2}$, $z$ is the perpendicular distance between the Chern insulator and the atom, $c$ is the speed of light, and $r_{ss}$, $r_{sp}$, $r_{ps}$ and $r_{pp}$ are reflection coefficients. The formulae for the reflection coefficients are derived in Ref.~\cite{lu2022}, and we refer the interested reader to that paper for details. The reflection coefficients depend on the conductivity tensor of the Chern insulator. In order to account for the effect of the frequency dispersion on the Casimir-Polder interaction, we apply Kubo's formula for the conductivity tensor and adopt the Qi-Wu-Zhang model of the Chern insulator, which is described by the band Hamiltonian
\be 
H = t \sin (k_x a) \sigma_x + t \sin (k_y a) \sigma_y + (t \cos (k_xa) + t \cos (k_y a) + u) \sigma_z,
\label{eq:QWZ}
\ee 
where $t$ is the hopping strength, $u$ is related to the band gap, $\sigma_x$, $\sigma_y$ and $\sigma_z$ are Pauli matrices, and $a$ is the lattice constant. In what follows, we set $u = t$. The behavior of the conductivity tensor can be broadly divided into three qualitatively distinct regimes depending on the photon's frequency~\cite{lu2020} (see Figure~\ref{dispersion}). For $\omega < 2t/\hbar$ (low-frequency regime), the frequency is not large enough to cause valence electrons to be excited to the conduction band, and the real part of the longitudinal conductance of the Chern insulator is zero. For $2t/\hbar \leq \omega \leq 6t/\hbar$ (intermediate-frequency regime), the valence electrons can be excited onto the conduction band, and the real part of the longitudinal conductance is nonzero. For $\omega > 6t/\hbar$ (high-frequency regime), the medium becomes effectively transparent. 

\

The resonant contribution $\delta E_1^{{\rm cl}}$ is given by 
\be
\delta E_1^{{\rm cl}}=- 
\frac{1}{2} \mu_a^{10}  
\big( G^R_{ab}({\bf r}_0,{\bf r}_0; \omega_{10}) 
+ G_{ba}^{R\ast}({\bf r}_0,{\bf r}_0; \omega_{10}) \big) \mu_b^{01}. 
\ee
For the lower-energy state, the Casimir-Polder shift, $\delta E_0^{CP}$, involves only the nonresonant contribution: 
\be
\delta E_0^{CP} = \delta E_0^{{\rm vdw}} = \frac{1}{\pi} \mu_a^{01} \mu_b^{10}  \,
\int_0^\infty \!\!\!\! d\xi \, 
\Bigg(
\frac{\omega_{01} \big( G^R_{ab}({\bf r}_0,{\bf r}_0; i\xi) + G_{ba}^{R}({\bf r}_0,{\bf r}_0; i\xi) \big)}{2(\omega_{01}^2 + \xi^2)}
-
\frac{\xi \big( G^R_{ab}({\bf r}_0,{\bf r}_0; i\xi) - G_{ba}^{R}({\bf r}_0,{\bf r}_0; i\xi) \big)}{2i(\omega_{01}^2 + \xi^2)}
\Bigg).  
\ee
From the above expressions for the nonresonant Casimir-Polder shift $\delta E_0^{{\rm vdw}}$, there is a term proportional to the antisymmetric part of the Green tensor, which would be zero for a reciprocal material, but nonzero for a nonreciprocal material. In the case of a Chern insulator, the antisymmetric part is nonzero, and as we shall see subsequently, leads to the possibility of a repulsive nonresonant Casimir-Polder force. 

\section{Resonant Casimir-Polder shift on an atom near a dispersive Chern insulator}
\label{sec:CPshift-res}

\begin{figure}[h]
    \centering
      \includegraphics[width=\textwidth]{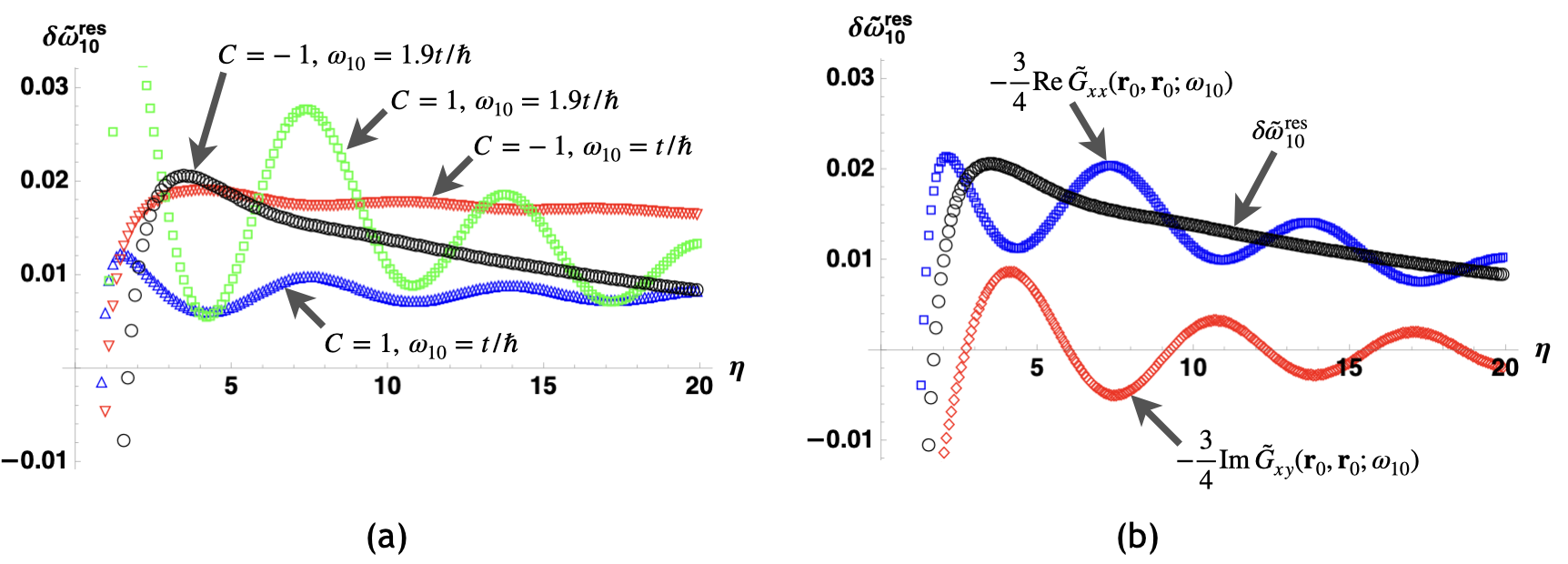}
      \caption{(a)~Behavior of the non-dimensionalized resonant Casimir-Polder shift $\delta\widetilde{\omega}_{10}^{\rm res} \equiv \delta E_1^{\rm cl} / \hbar R_{10}^{(0)}$ (circles) as a function of the non-dimensionalized distance $\eta \equiv 2\omega_{10}z_0/c$ (where $R_{10}^{(0)} \equiv 4\omega_{10}^3 \mu^2/3 \hbar c^3$ is the unperturbed transition rate), for a right circularly polarized dipole in front of a $C = 1$ Chern insulator with $\omega_{10}=t/\hbar$ (up triangles) and $\omega_{10}=1.9t/\hbar$ (squares), and for 
a right circularly polarized dipole in front of a $C = -1$ Chern insulator with $\omega_{10}=t/\hbar$ (down triangles) and $\omega_{10}=1.9t/\hbar$ (circles). 
      (b)~Behavior of $\delta\widetilde{\omega}_{10}^{\rm res}$ as a function of $\eta$ for a configuration consisting of a right circularly polarized dipole in front of a $C = -1$ Chern insulator, 
      with $\omega_{10} = 1.9t/\hbar$ (circles). Also shown are the behaviors of $-(3/4){\rm Re} \, \tilde{G}_{xx} \equiv - (3/4)(c/\omega_{10})^3 {\rm Re} \, G_{xx}$ (squares) and $-(3/4){\rm Im} \, \tilde{G}_{xy} \equiv - (3/4)(c/\omega_{10})^3 {\rm Im} \, G_{xy}$ (diamonds), which contribute to $\delta\widetilde{\omega}_{10}^{\rm res}$ for this configuration.} 
     \label{monotonic-decay}
  \end{figure}

To bring out the salient features of the Casimir-Polder behavior which are present for an atom near a Chern insulator but absent for an atom near a reciprocal material, we shall focus on the case of an excited atom in a right circularly polarized state with the quantization axis perpendicular to the plane of the Chern insulator. An example is a Sodium atom in the ${\rm 3p} \, ^2P_{3/2}$ state polarized to $m_J=3/2$, which can de-excite to the state ${\rm 3s} \, ^2S_{1/2}$ via an electric dipole transition. In this case, ${\bm{\mu}}^{10} = \mu(1,i,0)/\sqrt{2}$, and the resonant Casimir-Polder shift is given by 
\be
\delta E_1^{{\rm cl}}  
= 
- \mu^2 
\Big( {{\rm Re}}\, G_{xx}^R(\rv_0,\rv_0; \omega_{10}) 
+ {{\rm Im}}\, G_{xy}^R(\rv_0,\rv_0; \omega_{10}) \Big). 
\label{E1cl}
\ee
For the case of a right circularly polarized dipole and a $C=1$ Chern insulator, $\delta E_1^{{\rm cl}}$ typically undergoes an oscillatory decay with atom-surface separation (Figure~\ref{monotonic-decay}(a)), with the oscillations stemming from the presence of an exponential phase factor in the reflection Green tensor (Equation~(\ref{GR0})). On the other hand, for the same dipole in front of a $C=-1$ Chern insulator in the low-frequency regime, the oscillations of $\delta E_1^{{\rm cl}}$ can be strongly reduced (as we see in Figure~\ref{monotonic-decay}(a) for $\omega_{10} = t/\hbar$), or even disappear altogether, leading to a monotonic and non-oscillatory decay, as we see in the same figure for $\omega_{10} = 1.9t/\hbar$. The decay becomes monotonic as the second term on the right-hand side of Equation~(\ref{E1cl}) (i.e., $-(3/4){\rm Im} \, G_{xy}$) acquires a negative sign and thus oscillates anti-phasally to the first term (i.e., $-(3/4){\rm Re} \, G_{xx}$), causing the oscillatory behaviors of the two terms to cancel out (Figure~\ref{monotonic-decay}(b)). Correspondingly, the resonant Casimir-Polder force becomes monotically repulsive over a large range of atom-surface separations. 

\section{Nonresonant Casimir-Polder repulsion on an atom near a nondispersive Chern insulator}
\label{sec:CPshift-nonres}

Lastly, we consider the nonresonant Casimir-Polder shift of an atom near a Chern insulator. 
For the lower- and upper-energy states, let's consider the 5s5p $^3P_2$ state polarized to $m_J = 2$ of Strontium-88 (which can be prepared by optical pumping) and the 5s6s $^3S_1$ state respectively~\cite{footnote-david}, with the electric dipole transition again given by ${\bm{\mu}}^{10} = \mu(1,i,0)/\sqrt{2}$.
The lower-energy state is metastable, with a lifetime of the order of seconds, which means we can effectively regard the metastable state as a ground state within the timeframe set by its lifetime. 
Since the reflection Green tensor evaluated at imaginary frequencies is real-valued, the nonresonant Casimir-Polder shift of the metastable atomic state is given by 
\be
\delta E_0^{{\rm vdw}}
=
- \frac{\mu^2}{\pi} 
\int_0^\infty \!\!\!\!\! \ d\xi \,\,
  \left(
  \frac{\omega_{10} \, G_{xx}^R(\rv_0,\rv_0; i\xi)}{\omega_{10}^2 + \xi^2}  
  + \frac{\xi \, G_{xy}^R(\rv_0,\rv_0; i\xi)}{\omega_{10}^2 + \xi^2} 
  \right), 
\label{Evdw_0}
\ee
For our purpose, we consider a nondispersive Chern insulator. Firstly, this leads to a relatively simple analytical result for the far-field nonresonant Casimir-Polder force. Secondly, at large atom-surface separations, the nonresonant Casimir-Polder interaction is dominated by modes with small frequencies, and to a first approximation, we can replace the Chern insulator by one with a nondispersive conductivity, for which the longitudinal conductance vanishes and the Hall conductance is $Ce^2/h$. Correspondingly, the reflection coefficients are given by 
\be
r_{ss} = -r_{pp} = - \frac{(C\alpha)^2}{1+(C\alpha)^2}, \quad
r_{sp} = r_{ps} = - \frac{C\alpha}{1+(C\alpha)^2}, 
\ee
where $\alpha = e^2/\hbar c$ is the fine-structure constant. We obtain for the Green tensor components 
\begin{subequations}
\ba
G^R_{xx}({\bf r}_0, {\bf r}_0, i\xi) 
&=& 
G^R_{yy}({\bf r}_0, {\bf r}_0, i\xi) 
=
\frac{(C\alpha)^2}{1+(C\alpha)^2} \left( \frac{\xi^2}{4c^2 z_0} e^{-2\xi z_0/c} + \frac{\Gamma(3, 2\xi z_0/c)}{16z_0^3} \right), 
\\
G^R_{xy}({\bf r}_0, {\bf r}_0, i\xi) 
&=& 
- G^R_{yx}({\bf r}_0, {\bf r}_0, i\xi) 
=
\frac{C\alpha}{1+(C\alpha)^2} \left( \frac{\xi^2}{2c^2  z_0} + \frac{\xi}{4c z_0^2} \right) e^{-2\xi z_0/c}.  
\label{GRxy-non}
\ea
\end{subequations}
Here, $\Gamma(3,x)$ is an incomplete gamma function, defined by $\Gamma(a,x) = \int_x^\infty t^{a-1} e^{-t} dt$, whose leading-order behavior in the asymptotic expansion for large $x$ is given by $\Gamma(3,x) \approx (x^2 + 2x + 2) e^{-x}$~\cite{footnote-gamma}. 
In the far-field limit, this leads to 
\be
G^R_{xx}({\bf r}_0, {\bf r}_0, i\xi) 
=
G^R_{yy}({\bf r}_0, {\bf r}_0, i\xi) 
\approx
\frac{(C\alpha)^2}{1+(C\alpha)^2} 
\left( \frac{\xi^2}{2 c^2 z_0} + \frac{\xi}{4 c z_0^2} + \frac{1}{8 z_0^3} \right) 
e^{-2\xi z_0/c}.  
\label{GRxx-non}
\ee
Substituting Equations~(\ref{GRxy-non}) and (\ref{GRxx-non}) into Equation~(\ref{Evdw_0}) yields 
\ba
\delta E_0^{{\rm vdw}} &=& 
- \frac{\mu^2}{\pi} 
\frac{C\alpha}{1+(C\alpha)^2} 
\int_0^\infty \!\!\! d\xi \, 
e^{-2\xi z_0/c}
\bigg(
\frac{C\alpha \,  \omega_{10}}{\omega_{10}^2+\xi^2} 
\left( 
\frac{\xi^2}{2 c^2 z_0} + \frac{\xi}{4 c z_0^2} + \frac{1}{8 z_0^3} 
\right)
\nonumber\\
&&+
\frac{\xi}{\omega_{10}^2 +\xi^2}
\left(
\frac{\xi^2}{2c^2  z_0} + \frac{\xi}{4c z_0^2}
\right)
\bigg)
\ea
For large $z_0$, the frequency integral is dominated by modes with small $\xi$, and we can approximate $\omega_{10}^2+\xi^2$ by $\omega_{10}^2$ in the denominator. Since $(\xi/c)e^{-2\xi z_0/c} = -(1/2)\partial e^{-2\xi z_0/c}/\partial z_0$, we have 
\ba
\delta E_0^{{\rm vdw}} 
&\approx&
- \frac{\mu^2}{\pi}
\frac{C\alpha}{1 + (C\alpha)^2}
 \int_0^\infty \!\!\! d\xi 
\bigg(
\frac{C\alpha}{\omega_{10}} 
\left(
\frac{1}{8z_0} \frac{\partial^2}{\partial z_0^2} 
- \frac{1}{8z_0^2} \frac{\partial}{\partial z_0} 
+ \frac{1}{8 z_0^3}
\right) 
e^{-2\xi z_0/c} 
\nonumber\\
&&+ 
\frac{c}{\omega_{10}^2}
\left(
- \frac{1}{16 z_0} \frac{\partial^3}{\partial z_0^3} 
+ \frac{1}{16 z_0^2} \frac{\partial^2}{\partial z_0^2} 
\right) 
e^{-2\xi z_0/c} 
\bigg). 
\ea
By interchanging the order of integration and differentiation, we obtain  
\be
\delta E_0^{{\rm vdw}} 
=
-\frac{\mu^2 c}{4\pi \omega_{10}} 
\frac{C\alpha}{1 + (C\alpha)^2}
\left(
\frac{C\alpha}{z_0^4} + \frac{c}{\omega_{10} z_0^5}
\right). 
\ee
From the above, we obtain the nonresonant Casimir-Polder force acting on the metastable state:
\be
f^{{\rm vdw}} = - \frac{\partial \delta E_0^{{\rm vdw}}}{\partial z_0} =  
-\frac{\mu^2 c}{4\pi \omega_{10}} 
\frac{C\alpha}{1 + (C\alpha)^2}
\left(
\frac{4 C\alpha}{z_0^5} + \frac{5 c}{\omega_{10} z_0^6}
\right). 
\label{fCP-nonres-far}
\ee
In the limit that $z_0 \to \infty$ the first term dominates and the force is attractive. 
We see that this leading-order term is proportional to $(C\alpha)^2/(1+(C\alpha)^2)$, which suggests that one could perform a Casimir-Polder force measurement on the atom in the far-field region to determine the magnitude of the Chern number. 
Moreover, for values of $C$ such that $|C\alpha| < 1$, there is a window of separations in which the second term in Eq.~(\ref{fCP-nonres-far}) has a larger magnitude than the first term. If $C = -1$, the second term is positive, and the nonresonant Casimir-Polder force is repulsive for 
\be
z_0 < \frac{5c}{4|C|\alpha \omega_{10}}. 
\ee
We note that this range is obtained for the far-field region, where $z_0 > c/\omega_{10}$.  
If we take the lower- and higher-energy states of the two-level system to be the $m_J = 2$ polarized 5s5p $^3P_2$ state and the 5s6s $^3S_1$ state of Strontium-88 respectively, the electric dipole transition wavelength is 707.202 nm, and the range of $z_0$ for nonresonant Casimir-Polder repulsion is given by $1.1 \times 10^{-7} \, {\rm m} < z_0 < 1.9 \times 10^{-5} \, {\rm m}$.

\section{Discussion and Conclusion} 

We have considered the behaviors of the resonant and nonresonant Casimir-Polder interaction between a two-state atomic system and a Chern insulator, where the atomic states are connected by an electric dipole transition. We found that for a right circularly polarized excited atomic state, the resonant Casimir-Polder energy shift can monotonically decay over a large range of atom-surface separations if $C=-1$ and the ratio of the atomic transition energy to the hopping strength is sufficiently small, leading to a monotonic repulsive resonant Casimir-Polder force over the same separation range. On the other hand, the nonresonant Casimir-Polder force acting on a right circularly polarized metastable atomic state in front of a $C=-1$ Chern insulator is expected to become repulsive for separations greater than $c/\omega_{10}$ but smaller than $5c/ 4 |C| \alpha \omega_{10}$ (where $\alpha \approx 1/137$ is the fine-structure constant and $c$ is the speed of light), whereas the nonresonant Casimir-Polder force acting on the same atomic state in front of a $C=1$ Chern insulator is expected to be attractive at all separations. For both $C=1$ and $C=-1$, the nonresonant Casimir-Polder force is predicted to become attractive at sufficiently large separations and proportional to $(C\alpha)^2/(1+(C\alpha)^2)$, which evinces a topological signature of the Chern insulator. These two features of the nonresonant Casimir-Polder force can in principle be used to identify both the magnitude and sign of the Chern number of the Chern insulator. 

\acknowledgments 

The author would like to thank the organizers of the Fifth International Symposium on the Casimir Effect for the opportunity to present this paper, and also thanks Professors Martial Ducloy and David Wilkowski for constructive discussions.

\end{document}